\documentstyle[12pt,plenum,epsf]{book}
\begin{document}
%
%
\chapter{NONEQUILIBRIUM EFFECTS IN THE TUNNELING
CONDUCTANCE SPECTRA OF SMALL METALLIC PARTICLES}
\author{Oded Agam}
\affiliation{The Racah Institute of Physics\\
The Hebrew University\\
Jerusalem, 91904\\
Israel}
\section{1~~Introduction}

Trace formulae and the non-linear supersymmetric 
$\sigma$-model are basic analytical tools used successfully 
in the fields of quantum chaos and disordered systems. Both are designed to
treat systems with a small number of degrees of freedom. 
Hence they are limited in their possibility of analyzing 
many-body systems where interparticle interactions play an 
important role, and the number of degrees of freedom is large.

On the other hand, many experimental studies of
quantum chaos use systems which consist in a large number
of interacting particles, for example quantum dots or disordered
metallic particles. Having an elaborate single-particle
description of these systems, it is of prominent importance
to understand the role of interactions,
the range of applicability of a single-particle picture, and the interplay
between chaos and interparticle interactions. 

In this respect, an important observation is that strong chaotic 
dynamics, on the level of  non-interacting single-particle description, 
provides us with the possibility of analyzing interacting many-body 
systems by a systematic perturbative approach. The
small parameter of this perturbation theory is $1/g$, where 
$g=t_H/t_c$ is the dimensionless conductance, i.e. the ratio of the Heisenberg 
time, $t_H$ (the inverse mean level spacing), to the classical 
relaxation time, $t_c$.  

The general form of the interaction Hamiltonian 
in which  particles interact via a
two-body potential $U({\bf r}, {\bf r'})$ is
\begin{equation}
H_{\mbox{int}} = \frac{1}{2} \sum_{ ijkl} \sum_{ \sigma \sigma'} U_{ijkl}
c^\dagger_{i\sigma} c^\dagger_{j \sigma'} c_{k \sigma'} c_{l \sigma},
\label{Hint}
\end{equation}
where $c^\dagger_{i \sigma}$ and $c_{i\sigma}$ are  
the creation and annihilation operators for a particle in state 
$\psi_i$ and spin $\sigma$, while
\begin{eqnarray}
U_{ijkl} = \int d {\bf r} d {\bf r'} 
U({\bf r}, {\bf r'}) \psi_i^*({\bf r})
\psi_j^*({\bf r'})\psi_l({\bf r})\psi_k({\bf r'}) \nonumber
\end{eqnarray}
are the matrix elements of the interaction potential.
These matrix elements can be divided into 
two groups according to their typical magnitude.
One contains diagonal matrix elements, namely those $U_{ijkl}$
in which two pairs of indices are identical.
All the other matrix elements, which we call off-diagonal,
are included in the second group. In appendix A it is shown that
the typical magnitude of  off-diagonal matrix elements 
is as small as $d/g$ where $d$ is the single-particle mean level spacing
and $g$ is the dimensionless conductance.
The same smallness restricts also the fluctuations
in the diagonal matrix elements. Therefore,
the interaction Hamiltonian of electrons in a quantum dot
takes the form
\begin{equation}
H_{\mbox{int}} = \frac{e^2}{2C} \left( 
\sum_{i \sigma} c^\dagger_{i \sigma } c_{i \sigma} - N_0 \right)^2
- \lambda \sum_{i,j} c^\dagger_{i\uparrow} c^\dagger_{i \downarrow} 
c_{j \uparrow} c_{j\downarrow} + \frac{a}{2} \! \sum_{i j, \sigma \sigma'}
 c^\dagger_{i\sigma} c^\dagger_{j \sigma'} c_{i \sigma'} 
c_{j \sigma} + O(d/g). \label{MFH}
\end{equation}
The first term of this formula is the orthodox model
\refnote{\cite{Averin91,review}}
representing the charging energy of the dot: $C$ is the total
capacitance of the dot, $N=\sum_{i \sigma} c^\dagger_{i \sigma } 
c_{i \sigma}$ is the total number of electrons in the dot, and
$N_0$ is a continuous parameter controlled by the gate voltage.
The second term, when $\lambda > 0$, represents an attractive interaction
which drives the grain into a superconducting state at sufficiently
low temperatures and weak magnetic field. The last term represents
the electron-electron interaction in the spin channel, 
and the coefficient $a$ is of order the single-particle mean 
level spacing, $d$.

Strong chaotic dynamics of the noninteracting particles
implies that 
\linebreak
$g \gg 1$. Consequently, all off-diagonal
matrix elements $U_{ijkl}$ are proportional to $1/g$, and the  mean field 
approximation (\ref{MFH}) for the interacting Hamiltonian (\ref{Hint})
is justified.  Indeed, many phenomena of normal and superconducting 
metallic grains are described by the approximation
(\ref{MFH}). The most prominent one is the Coulomb blockade, which
is essentially the quantization of the number of electrons
in the grain away from the charge  degeneracy point. 
Because of this quantization, the zero-bias
conductance of the system vanishes, while the current $I$ as the function
of the source-drain voltage $V$ shows a threshold behavior. The fine
structure of the current--voltage curve is associated with the 
single-electron levels of the system\refnote{\cite{AverinKorotkov,Leo}}.

Nevertheless, there are interesting phenomena emerging from
the fluctuations of the interaction matrix elements, i.e.
with the $O(d/g)$ corrections to (\ref{MFH}). 
In this review we analyze two experiments of
Ralph, Black and Tinkham\refnote{\cite{RBT, Ralph1}} and show that
fluctuations of the interaction energy, although small as $d/g$, 
clearly manifest themselves in the differential conductance spectra
of ultrasmall metallic grains. The small 
effects of fluctuations in the charging energy 
are especially pronounced due to the
fact that the system is driven out of equilibrium, and is
able to explore several high excited states at relatively low
source-drain voltage.
  
The experimental system consists of a single aluminum particle 
connected to external leads via high resistance (1 -- 5 M$\Omega$) 
tunnel junctions formed by oxidizing the surface of the particle.
\begin{figure}
\begin{center}
\leavevmode
\epsfxsize=10.0cm
\epsfbox{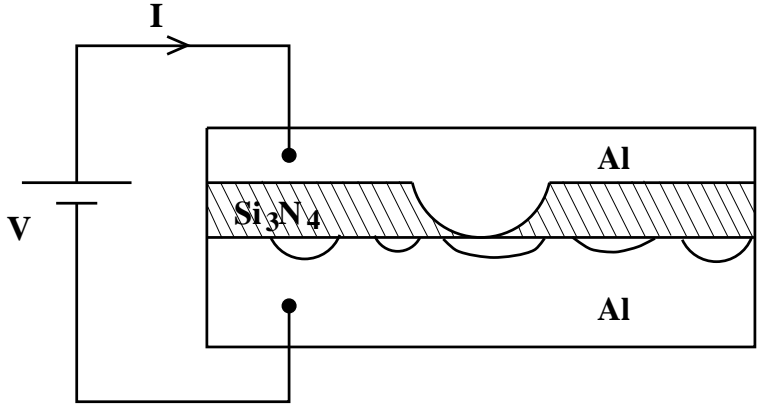}
\end{center}
\caption{A schematic illustration of the device used in the
experiments of Ralph, Black and Tinkham for measuring the
differential conductance spectra of ultrasmall aluminum grains.}
\end{figure}
The device (see illustration in Fig.~1) 
is fabricated using electron beam lithography and reactive
ion etching to form a bowl-shaped hole in an insulating
Si$_3$N$_4$ membrane. The opening  at the lower edge of the
membrane is 3-10 nm in diameter. Al is evaporated onto the bowl-shaped
side of the membrane, and subsequently the Al surface is oxidized.
The oxide layer forms a tunnel barrier in the vicinity of the small
hole in the membrane. The membrane is then flipped up side down
and a small amount of
Al is deposited. Because of surface tension, the Al forms a layer of
electrically isolated  particles, a few nanometers in size. 
Following a 
second oxidation, a thick  layer of Al is deposited on top of the 
particles.\footnote{
In other configurations of these devices a gate electrode of ring shape
is deposited after flipping the membrane, and  the same 
procedure follows the oxidation of the gate.}
In approximately 25\% of the devices one Al particle covers
the hole in the nitride membrane, so that the electrons passing 
between the leads tunnel trough the metal particle.

The capacitances and the resistances of the tunnel junctions are 
estimated  by fitting the large scale $I-V$ curves, $eV \sim e^2/C$, 
to the Coulomb blockade
staircase pattern. From the capacitances one can determine the area
of the tunnel junctions and the volume of the grain
which is used in turn to estimate the single-particle 
mean level spacing\refnote{\cite{RBT}}.
\begin{figure}
\begin{center}
 \leavevmode
    \epsfxsize=11cm
\epsfbox{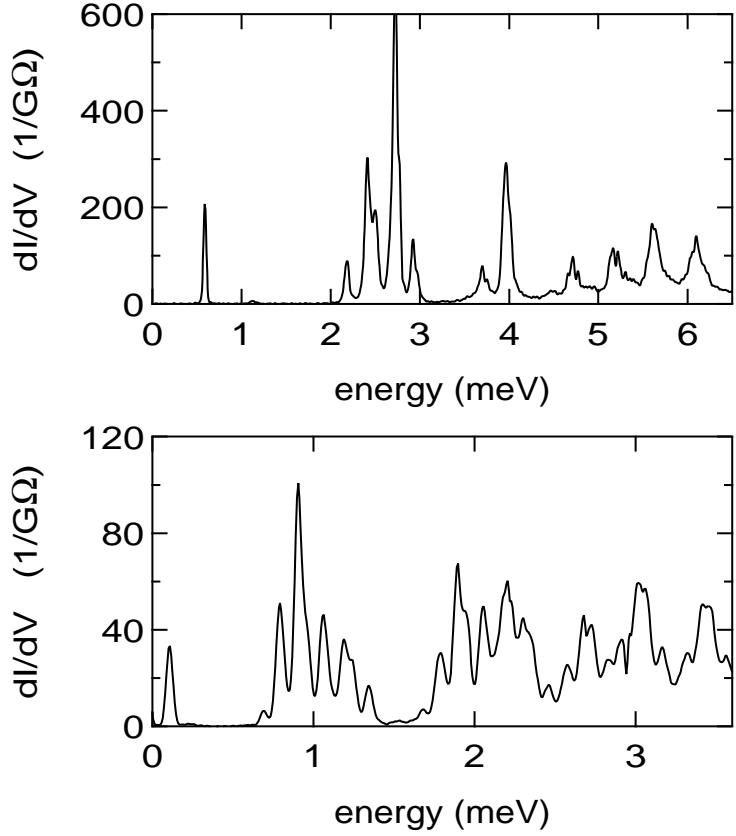} 
 \end{center}
\caption{The low temperature (30 mK) differential conductance
$dI/dV$ versus bias energy of ultrasmall Al particles with volumes 
 $\approx$ 40 nm$^3$ (upper panel) $\approx$ 100 nm$^3$ (lower panel).
The first resonance is isolated while subsequent
resonances are clustered in groups. The distance between
nearby groups of resonances is approximately the 
single-particle mean level spacing $d$. (From  Ref.~[5]).}
\end{figure}

In this review we focus our attention on scales of the $I-V$ curves
which are much smaller than that of the Coulomb blockade,
scales over which the single-particle mean level spacing, $d$, 
and the fluctuation in the charging energy, $d/g$, are resolved.  
Fig.~2 displays the differential conductance, $dI/dV$, of  
two different {\em normal} metallic  particles 
(of sizes roughly 2.5 and 4.5 nm) as a function of 
the source--drain bias energy $eV$. The spectra display three clear features: 
\begin{enumerate}
\item
The low resonances of the differential conductance
 are grouped in clusters. The distance between 
nearby clusters is of order the mean level spacing $d$ of the 
noninteracting electrons in the dot. 
\item The first cluster contains
only a single resonance. 
\item
Higher clusters consist of several 
resonances spaced much more closely than $d$.
\end{enumerate}
In section 2 it will be shown that these features are manifestations
of the interplay between electron-electron interactions and 
nonequilibrium effects\refnote{\cite{nonequilibrium}}. 
Each cluster of resonances is identified 
with one excited single-electron state, and each resonance in turn
is associated with  a different occupancy configuration of the 
metal particle's other single-electron states. The appearance of 
multiple resonances reflects the strongly nonequilibrium state of
the particle. 
\begin{figure}
  \begin{center}
 \leavevmode
    \epsfxsize=10 cm
 \epsfbox{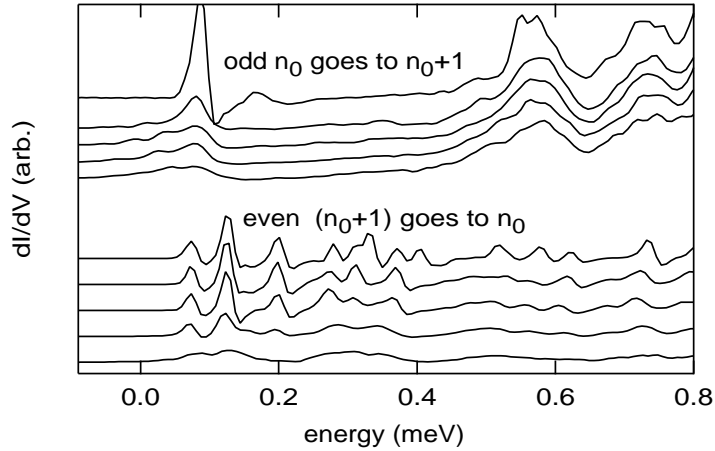} 
 \end{center}
\caption{The tunneling resonances of superconducting grains
in the odd (upper scans) and the even (lower scans)
charging states. Different scans correspond to different value of
the gate voltage, and are artificially shifted in energy to
align peaks due to the same eigenstate. In contrast with Fig.~2
the first resonance, in the odd charging state, develops a substructure
when shifted by the gate voltage. (From Ref.~[6]).}
\end{figure}

In another experiment\refnote{\cite{Ralph1}}, 
Ralph, Black and Tinkham measured the tunneling
resonance spectra of ultrasmall {\em superconducting} grains. The number of
electrons in the system was controlled by a gate voltage. The results
of this experiment, depicted in Fig.~3, show that:
\begin{enumerate}
\item For the ground 
state of the grain with an even number of electrons,  
the first peak of the differential conductance is merely shifted by 
the gate voltage $V_g$. The shape of this peak does not change over
a large  interval of $V_g$. Contrarily, if the grain contains an odd 
number of electrons,  the height of the first peak rapidly reduces 
with a change of the gate  voltage, and a structure of 
subresonances develops  on the low-voltage shoulder of this peak.
\item
The characteristic energy scale between subresonances of the first peak 
is of the order of the mean level spacing $d$.
\end{enumerate}

These observations contrast the results for the
normal case in which the first peak did not split.
Nevertheless, it was suggested in Ref.~[\cite{Ralph1}] that
the substructure of the first peak is still associated with
nonequilibrium steady states of the grain.  In section 3, the origin 
of these nonequilibrium states and the mechanism which generates them will 
be clarified\refnote{\cite{AA}}. 

The explanations for both experiments discussed here
rely on the assumption that the systems are stimulated into 
steady states which are far from equilibrium, namely that
relaxation processes are too slow to maintain
the system in equilibrium. In section 4 we summarize the 
various relaxation processes and estimate the inelastic
time, $\tau_{in}$, for electrons in the dot. 
The results will be summarized in section 5. 

\section{2~~Normal grains}

Our model for the experimental system is given by 
the Hamiltonian: $H=H_0+H_{T}+H_{\mbox{int}}$. Here $H_0$ describes the 
noninteracting electrons in the left (L) and right (R) 
leads and in the metallic grain\footnote{Unless explicitly written,
from now on single-particle and spin states will be denoted by 
a single subscript.},
\begin{equation}
H_0= \sum_{\alpha =L,R} \sum_q \xi_{\alpha q} 
d^\dagger_{\alpha q} d_{\alpha q}+ \sum_l \xi_l c^\dagger_l c_l. \label{H0}
\end{equation}
Tunneling across the barriers is described by 
\begin{equation}
H_{T}= 
\sum_{\alpha =L,R} \sum_{q,l} T^{(\alpha)}_{ql}d^\dagger_{\alpha q}c_l +
 \mbox{ H.c.}, \label{Ht}
\end{equation}
where $T^{(\alpha)}_{ql}$ are the tunneling matrix elements. Interaction
effects given by (\ref{Hint}) are taken into account only for the 
electrons in the grain, but including screening by image 
charges in the leads. 

For the ultrasmall
aluminum grains considered here, one can neglect 
superconducting pairing since the single-particle mean level spacing,
$\approx \! 1$ meV, is larger than the BCS superconducting gap
which is $0.18$  meV\refnote{\cite{Delft96}}.   
Under this condition, the interaction term of the electrons is 
generally approximated by the orthodox model, $H_{\mbox{int}} \approx
(e\sum_l c^\dagger_l c_l )^2/C$, where $C$ is the effective 
capacitance of the grain\refnote{\cite{Averin91}}.
Within this approximation the charging energy depends only on 
the total number of electrons in  the dot, but not on their 
particular occupancy configuration. 
The orthodox model is able to account for the
Coulomb blockade\refnote{\cite{Averin91}}, and the Coulomb staircase behavior 
of the current as the  number of extra tunneling electrons in the dot
increases. It can also be generalized to describe features
on the scale of the single-particle level spacing\refnote{\cite{Averin90}}.
However, the orthodox model cannot account for the clusters
of resonances in Fig.~2, since these result from fluctuations, 
$\delta U$, in the interaction energy between pairs of electrons.

We focus our attention on the (experimental) voltage regime  
where there is no more than one extra tunneling electron in the dot. 
At small voltage bias, $V$, within the Coulomb-blockade regime 
[Fig.~4(a)], current does not flow through the system. 
Current first starts to flow when one state $i$ inside the grain
becomes available for tunneling through the left barrier, say, 
as illustrated in Fig.~4(b). As the system becomes charged 
with an additional electron, the potential energy of
the other electrons in the dot increases by $U\simeq e^2/C$, 
and some of the lower energy occupied electronic states
are raised above the right lead chemical potential [in Fig.~4(b) 
these ``ghost'' states are shown as dashed lines].
\begin{figure}
  \begin{center}
 \leavevmode
    \epsfxsize=9cm
\epsfbox{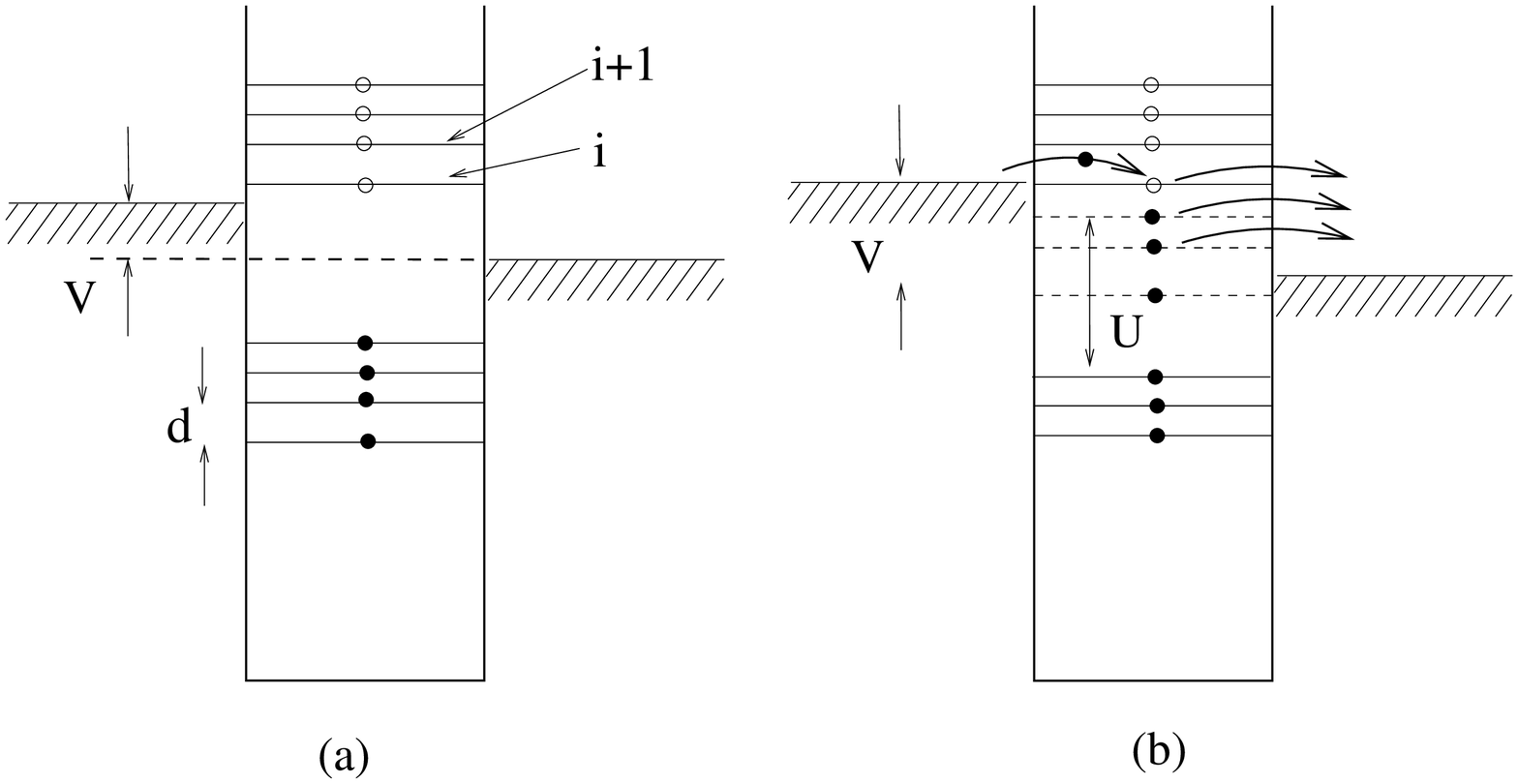}
 \end{center}
  \begin{center}
 \leavevmode
    \epsfxsize=9cm
\epsfbox{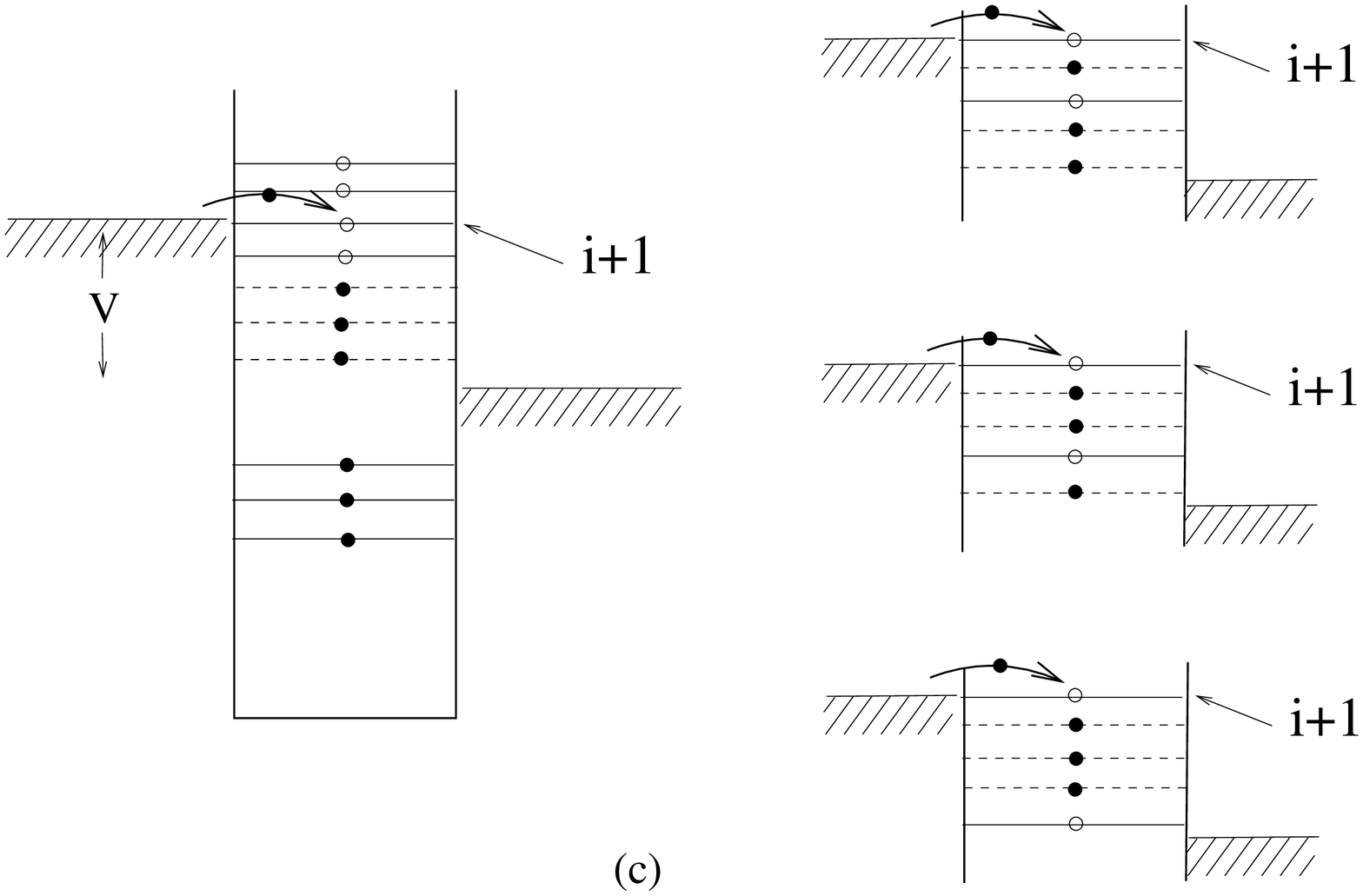}
 \end{center}
\caption{An illustration of transport through the metal particle
at various values of the source--drain voltage $V$. Filled single-particle 
levels are indicated by full circles and empty ones by open circles.  
$U$ is the charging energy, and $d$ is the single-particle 
mean level spacing. (a) The system at small voltage bias within the 
Coulomb blockade regime; (b) $V$ corresponding to the first resonance
in Fig.~2. The thin dashed lines
indicate the energy of a level after an electron has tunneled
into the dot;  (c) $V$ near
the first cluster of resonances in Fig.~2. The
splitting within the first cluster originates from the 
sensitivity of level $i+1$ to the different possible
occupation configurations as shown.}
\end{figure}
Electrons can tunnel out from these states into the right lead
leaving the particle in an excited state.
There is, however, only one configuration of the electrons which allows 
an electron to tunnel into level $i$ 
from the left lead, namely all lower energy 
levels occupied. This implies that only a single resonance peak appears 
in the differential conductance at the onset of the current flow 
through the system (broken spin degeneracy would cause splitting 
of this peak).

The situation changes when $V$ increases such that 
electrons can tunnel from the left lead into the next higher 
available state $i+1$, as shown in Fig.4~(c). In this case, 
there are several possible occupancy configurations, 
on which the exact energy of level $i+1$ depends. 
The several possible energies of level $i+1$ lead to  
a cluster of resonances in the differential conductance 
of the grain. The scenario described above holds provided that 
inelastic processes are too slow to maintain equilibrium in the particle.

To explicitly demonstrate the splitting of resonances induced by 
fixed fluctuations in the interaction energy $\delta U$, 
model detailed-balance equations\refnote{\cite{Averin90}} were solved 
numerically and the corresponding differential conductance 
plotted in Fig.~5 by the solid line.
\begin{figure}
  \begin{center}
\leavevmode
    \epsfxsize=11.0cm
 \epsfbox{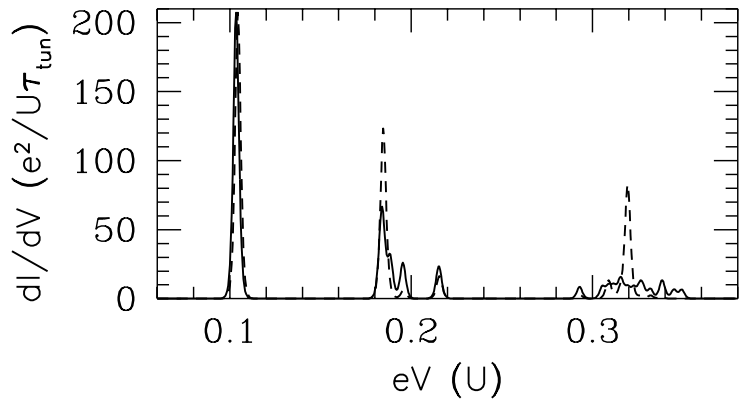}
\ 

\vspace{-2.2in}
\end{center}
 \caption{Model differential conductance obtained 
from nonequilibrium detailed-balance equations: 
solid line -- in the absence of inelastic processes, 
$1/\tau_{in}=0$; dashed line -- with inelastic
relaxation rate larger than the tunneling rate, 
$1/\tau_{in}=5/\tau_{tun}$.}
\end{figure}
The model system consists of 7 equally
spaced levels, occupied alternately by 4 or 5 electrons, 
in a current-carrying steady state. For simplicity, the tunneling 
rate into each level, $1/\tau_{tun}$ ($\Gamma_{L(R)}(\epsilon_l)$ in 
the notation of Ref.~[\cite{Averin90}]), 
is chosen to be uniform, and the voltage is applied by increasing
the left chemical potential. The temperature is 1\% of the 
mean level spacing $d$, and the variance of the 
fluctuations $\delta U$ in the interaction energy is $d/5$.   
In the absence of fluctuations ($\delta U=0$), $dI/dV$ consists of
single resonances spaced by $d$. 

To estimate the fluctuations in the interaction energy consider 
the Hartree term of the interaction energy, $U_H$. 
We wish to calculate the interaction energy difference associated 
with different occupation configurations of low energy states. 
Suppose that, as illustrated in Fig.~4(c), these  differ by a single 
occupation number, namely, in one configuration the state $j$ is empty 
and $j'$ is full while in the other $j'$ is empty and $j$ is full. 
Then 
\begin{eqnarray}
\delta U_H \!\!= \!\!\!\int \!\! dr_1 dr_2 |\psi_i({\bf r_1})|^2
U({\bf r_1},{\bf r_2})\! \left[ |\psi_{j'}({\bf r_2})|^2 \!\!-\!
|\psi_{j}({\bf r_2})|^2 \right], \nonumber 
\end{eqnarray}
where the index $i$ labels an electron state other than $j$ or $j'$, 
$U({\bf r_1},{\bf r_2})$ is the interaction potential.
Since wave functions of chaotic systems associated with different 
energies are statistically independent, $\langle \delta U_H \rangle =0$ 
where $\langle \cdots \rangle$
denotes ensemble or energy averaging. We are therefore interested in
fluctuations of $\delta U_H$ which emerge from the non-uniform 
probability distributions of the single-particle eigenstates
in real space. The calculation of $\langle \delta U_H^2 \rangle$
is similar to that presented in appendix A.  
The result for diffusive systems is
\begin{eqnarray}
\langle \delta U_H^2 \rangle = \left( c \frac{d}{g}\right)^2, \nonumber
\end{eqnarray}
where $c = \sqrt{2} \alpha \sum_{\bf n}|{\bf n}|^{-4} /\pi $ is a 
constant of order unity, and $\alpha$ equals two for system 

with time reversal symmetry and unity for systems without time reversal 
symmetry\footnote{This 
estimation does not take into account a  change in the 
potential due to the insertion of an additional electron. It was
argued that the latter effect may lead to an even stronger 
effect, i.e. $\delta U \sim d/\sqrt{g}$.\refnote{\cite{BMM}}}. 
The above estimate for the fluctuations in the charging energy
also applies for general chaotic systems, with $g= \hbar \gamma_1/d$ where 
$\gamma_1$ is the first non-vanishing Perron--Frobenius 
eigenvalue\refnote{\cite{Agam95}}, see appendix A.

The increase of the fluctuations in the interaction energy as
$g$ decreases is related to the fact that $g$ is a measure 
for the uniformity of the single-particle wave functions. The bigger
$g$ the more uniform are the wave functions and the less are the 
fluctuations in the interaction energy. 
Experimentally we find $g \approx 5$. Unfortunately, an analytical
estimate of $g$ requires precise knowledge of the shape and disorder
of the particle which we lack. A naive estimate of $g$ in ballistic 
systems is $\hbar/\tau d$, where $\tau$ is the time for an electron
at the Fermi energy to cross the system. The metallic 
grains of the experiment, however,  have a roughly pancake shape.
Assuming diffusive dynamics one can show that $g\tau d/\hbar 
\propto (z/r)^2$ where $z$ is the pancake thickness and $r$ is its radius.
$g$ is therefore much smaller than $\hbar/\tau d$.

When  $M$ available states below the highest
accessible energy level (including spin), are occupied by 
$M'<M$ electrons, there are ${\tiny (\! \! \begin{array}{c} 
M \\ M' \end{array}\! \!)}$ different occupancy configurations. 
The typical width of a cluster of resonances in this case is 
$W^{1/2} c d/g$ where $W=\mbox{min}(M-M', M')$. The width of 
a cluster of resonances therefore {\em increases} with the 
source--drain voltage. The distance between nearby peaks of the 
cluster, on the other hand, {\em decreases} as 
$W^{1/2}/ {\tiny (\! \! \begin{array}{c} M \\ M' \end{array}\! \!)}$. 
This behavior can be seen in Fig.~5.

\section{3~~Superconducting grains}

As illustrated by Fig.~4, the splitting of tunneling resonance 
peaks in normal metallic grains comes from the possibility 
of forming different occupation
configurations of single-particle states at sufficiently high
source-drain voltage. These configurations are reached by
resonant tunneling provided relaxation processes are sufficiently
slow. This picture  also explains the observation
that the first cluster in Fig.~2 contains only a single peak.
However, the data of Fig.~3 shows that 
the first peak in the tunneling resonance spectra  
also splits into several subresonances (see illustration in Fig.~6). 
This behavior appears when the superconducting grain contains an odd 
number of electrons and the gate voltage is such that
the dot is far from the charge degeneracy point.
\begin{figure}
\begin{center}
\leavevmode
\epsfxsize=8cm
\hspace*{-0.2cm}
\epsfbox{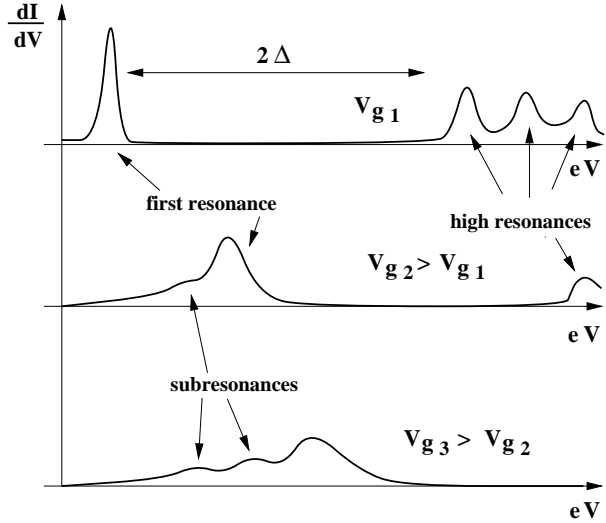} 
\end{center}
\caption{A schematic illustration of the differential conductance of
an ``odd'' superconducting grain as function of the source-drain
voltage $V$, at various gate voltages $V_g$. Higher resonances are
separated by the superconducting gap from the first one, and
subresonances are developed as the first resonance is shifted by the
gate voltage.}
\end{figure}

In this section we show that the development of a 
substructure in the first peak of the tunneling resonance
spectra is also associated with the generation of nonequilibrium
steady state. However in contrast with the resonant tunneling
mechanism used in the previous section for the high resonance 
peaks, here the nonequilibrium steady state is reached
by inelastic cotunneling processes.

The principal difference between odd and even grains is that 
all excitations of the latter are of energy larger than the superconducting
gap $2 \Delta$. Therefore, a source-drain voltage in the range $V < \Delta/e$ 
can not induce nonequilibrium states. Odd grains, on the other hand, 
contain one unpaired electron, which may be shifted to various 
single-electron levels with characteristic energy scale smaller than the
mean level spacing $d$. For this reason  even a small source-drain voltage
$d<eV<\Delta$ is sufficient to excite the grain.  
The mechanism of excitation is inelastic 
cotunneling\refnote{\cite{cotunneling}}. 
Tunneling into the excited grain requires less energetic 
electrons, and lead in turn to the substructure on the 
low-voltage shoulder of the of the first resonance, see Figs.~3 and 6. 
A closely related problem was considered by Averin and 
Nazarov\refnote{\cite{AverinNazarov}}, however, their theory assumed 
that relaxation processes prevent the formation of 
nonequilibrium states. As will be argued in the next section,
relaxation processes in ultrasmall metallic grains
are very slow, and therefore will be neglected in our theory.

To describe the effect quantitatively, 
we construct the master equations governing the time
evolution of probabilities of different electronic configurations of
superconducting grains allowing for second order cotunneling
processes. The solution of these equations for two limiting cases
(one in which two levels participate in the transport, 
and the other when a large number of levels contribute) 
explains the substructure of the first peak of the 
differential conductance illustrated in Fig.~6.

As in the previous section the model Hamiltonian consists of
three terms: $H=H_0+H_{T}+H_{\mbox{int}}$.
$H_0$, given by (\ref{H0}), describes the noninteracting electrons
in the leads and in the dot; $H_{T}$, given by (\ref{Ht}), is the 
tunneling Hamiltonian, and the interaction Hamiltonian
will be approximated by
\begin{equation}
H_{\mbox{int}}=\frac{e^2}{2C} (N-N_0)^2
- \lambda \sum_{i,j}  c^\dagger_{i\uparrow} 
c^\dagger_{i\downarrow} c_{j\uparrow} c_{j\downarrow}.
\label{interaction}
\end{equation}
where $N=\sum_{j,\sigma} c^\dagger_{j\sigma} c_{j\sigma}$ 
is the number of electrons in the dot,   
and $N_0$ is a continuous parameter controlled 
by the gate voltage.  $N_0$ determines the finite charging energy 
required to insert,  $U_+$, or to  remove, $U_-$, one electron, 
\begin{equation}
U_\pm=\frac{e^2}{C} \left[ \frac{1}{2}\pm (N-N_0) \right],\quad
|N-N_0| \leq \frac{1}{2}.
\label{U} 
\end{equation}

Consider the experimentally relevant case, $e^2/C \gg \Delta$, 
so that the grain has  well defined number of electrons. 
If this number is even $N=2m$, the ground state energy 
(we will omit charging part of the energy and restore it later), 
$E_{2m}= F_{2m}+2\mu m$, can be calculated in the
mean field approximation\refnote{\cite{textbook,Delft96}} 
by minimizing thermodynamic potential 
$F_{2m} = \sum_{k} (\xi_k - \epsilon_k) + \Delta^2/\lambda$
where $\epsilon_k= (\xi_j^2 + \Delta^2)^{1/2}$, with respect to
$\Delta$, and by fixing the chemical potential 
according to the number of electrons in the grain. All the
excited states of even dots are separated from the ground state 
by a large energy, $2\Delta$. 
Considering now the energy spectrum of an odd grain, $N=2m-1$,
we notice that the second term in Eq.~(\ref{interaction}) operates
only within spin singlet states. Therefore, to calculate the
low-lying excited states in this case, we fill the single-electron 
state $j$ with one electron, and then find the ground state of the 
remaining $2m$ electrons with state $j$ excluded from the 
Hilbert space. In the mean field approximation it corresponds 
to the minimization of the thermodynamic potential 
$F_{2m-1}^{(j)} = \sum_{k\neq j} (\xi_k -
\epsilon_k) + \Delta^2/\lambda + \xi_j$. The excited states with
energies smaller than $\Delta$ are characterized by a single
index, $j$ and will be denoted by $E_{2m-1}^{(j)}$.  
In what follows we will need the energy
cost of introducing an additional electron into the odd state: 
$U_+ +\varepsilon_j$, where $\varepsilon_j = E_{2m}-E_{2m-1}^{(j)}$. 
In appendix B it is shown that in the limit $\Delta \gg d$ the result is: 
\begin{equation}
\varepsilon_j = \mu_{2m} - \frac{3 d}{2} + \frac{\xi_j d}{2 \Delta}
 -\sqrt{ \xi_j^2 + \Delta^2}.
\label{epsilon}
\end{equation}

We turn now to the kinetics of a superconducting grain.
Consider the regime where $U_+ = U \leq
\Delta$, $U_- \approx \frac{e^2}{2 C} \gg U$,
and $\frac{e^2}{2 C} \gg \Delta \gg d$. 
We also assume  the conductance of the
tunnel barriers to be much smaller than  $e^2/h$,
and that the source-drain voltage is  small  $eV < \Delta$. 
The simplicity brought to the problem in this regime
of parameters stems from the fact that there is only one available
state with an even number of electrons (because $U_-\gg U_+$ 
one can only add an electron to grain but not subtract one), 
and whenever the grain contains an even number of electrons it is in its  
ground state. This imply that even grains
cannot be driven out of equilibrium state, while for odd grains
tunneling (and cotunneling) takes place via unique state. 

Henceforth, we concentrate on grains with an odd ground state. 
Let us denote by
$P_e$ the probability of finding the grain with an even number of
electrons, and by $P_j$ the probability to find the grain in the odd
state $j$.  Since these states are spin degenerate in the absence of
magnetic field, $P_j$ will denote the sum
$P_{j,\uparrow}+P_{j,\downarrow}$.  The master equations for the
probabilities $P_e, P_j$ have the form
\begin{eqnarray}
&&\frac{ d P_e}{ dt} =  \sum_j \left[ \Gamma_{o \to e}^{(j)} P_{j}
- 2\Gamma_{e \to o}^{(j)} P_e  \right], \label{rate-equations}  \\
&&\frac{d P_j}{dt} \!=\!2\! \sum_{i \neq j}\! \left[ 
\Gamma_{i \to j} P_i \!-\! \Gamma_{j \to i}
    P_j \right]\! +\!2 \Gamma_{e \to o}^{(j)} P_e \!-\! 
   \Gamma_{o \to e}^{(j)} P_j, \nonumber
\end{eqnarray}
where $\Gamma_{o \to e}^{(j)}$ and $\Gamma_{e \to o}^{(j)}$ are the
transitions rates from the odd $j$-th state to the even and from the
even to odd respectively, while $\Gamma_{i \to j}$ is the rate of
transition from the $i$-th to the $j$-th odd states.  Equations
(\ref{rate-equations}) are not independent, so they have to
be supplied with the normalization condition $P_e+ \sum_j P_j =1$.
Current in the steady state equals to the electron
flow through, say, the left barrier, and for positive $V$ it is given by
\begin{equation}
I =e \sum_j\left(\Gamma_{o\to e}^{(j)}+\Gamma_{j\to j}
 \right)P_j + 2e\sum_{j\neq i} \Gamma_{j\to i}^{(j)}P_j.
\label{current}
\end{equation}

Transition from
the $j$-th odd state into the even state occurs when $\mu_L >
U+ \varepsilon_j$. The amplitude of this transition is calculated
by first order perturbation theory in the tunneling Hamiltonian
(\ref{Ht}). Fermi's golden rule yields
\begin{equation}
\Gamma_{o \to e}^{(j)} = g_L \frac{ u_j^2 \rho_{Lj} d}{2 \pi \hbar}  
\theta(\mu_L-\varepsilon_j-U), 
\label{rate1}
\end{equation} 
where $g_L$ is the dimensionless conductance of the left tunnel
barrier per one spin,
$u_j = (1 + \xi_j/\epsilon_j)/2$ is the coherence factor, $\theta(x)$
is the unit step function, and
$\rho_{Lj}=\Omega |\psi_j(r_L)|^2$, where $\Omega$ is the volume of the 
grain and $\psi_j(r_L)$ is the value of $j$-th single-particle wave 
function at the left point contact $r_L$. Energies $\varepsilon_i$ are
given by Eq.~(\ref{epsilon}) and $U=U_+$ is defined in Eq.~(\ref{U}).
Similarly, the rate of transition from even state to $i$-th odd state,
by tunneling of an electron from the dot to the right lead, is
given by     
\begin{equation} 
\Gamma_{e \to o}^{(i)} = g_R \frac{v_i^2 \rho_{Ri} d}{ 2 \pi \hbar}  
\theta(U + \varepsilon_i-\mu_R), \label{rate2}
\end{equation}
where $g_R$ is the dimensionless conductance of the right tunnel barrier,
$v_i = (1 - \xi_i/\epsilon_i)/2$, and $\rho_{Ri}=\Omega |\psi_i(r_R)|^2$,
where $r_R$ is the position of the right point contact.

A change in the occupation configuration of the odd states occurs
via inelastic cotunneling\refnote{\cite{cotunneling}}. 
This mechanism is a virtual process in which
an electron tunnels into $j$-th available level and another electron
tunnels out from the $i$-th level. Calculating this rate by second order
perturbation theory in the tunneling Hamiltonian, one obtains
\begin{equation}
\Gamma_{j\to i} = \frac{g_L g_R d^2 u_j^2 v_i^2 \rho_{Lj} \rho_{Ri}
(eV-\varepsilon_j+\varepsilon_i) }{ 8 \pi^3 \hbar
(U + \varepsilon_j - \mu_L)(U + \varepsilon_i -\mu_R)}
\label{rate3}   
\end{equation}
for $eV>\varepsilon_j-\varepsilon_i$, $\mu_L < U + \varepsilon_j$,
$\mu_R < U + \varepsilon_i$, and zero otherwise.
$\Gamma_{j\to i}$ diverges in the limits $\mu_L \to U + \varepsilon_j$
and $\mu_R \to U + \varepsilon_i$. It signals that a real
transition takes over the virtual one. The region of
applicability of Eq.~(\ref{rate3}) is, therefore, $ U +
\varepsilon_j -\mu_L > \gamma$ and $ U + \varepsilon_i -\mu_R >
\gamma$ where $\gamma \sim gd/4\pi$ is the width of a single-particle
level in the dot due to the coupling to the leads, $g= g_L +
g_R$. However,
the interval of biases where Eq.~(\ref{rate3}) is not valid is narrow,
and to the leading approximation in $\hat{H}_T$ our results will be 
independent of this broadening.

Let us now apply Eqs.~(\ref{rate-equations}) and (\ref{current}) to
describe the appearance of the low-voltage substructure of the first
peak. We will consider two situations: (i) small voltage
such that only one subresonance can emerge on the shoulder of the leading one, 
and (ii) large voltage, $d\ll eV < \Delta$, where 
the substructure of the main resonance consists of 
a large number of subresonances.  

In the first case, the chemical potentials of the
left and right leads are such that transport through the grain
involves only two levels: $\varepsilon_0$ and $\varepsilon_{1}<\varepsilon_0$
corresponding to the ground and the first excited states of the odd
grain. We solve Eqs.~(\ref{rate-equations}) for probabilities $P_0,P_1$
and $P_e$ using Eqs.~(\ref{current}--\ref{rate3}). There are
two distinct regimes of the source-drain voltage: (1) 
$\mu_L < U + \varepsilon_0$ where transport is dominated by
cotunneling, and (2) $\mu_L \geq U + \varepsilon_0$
where state ``0'' is available for resonant tunneling.
The substructure of the first resonance in the differential conductance
appears in the first regime. Below we show that  
as  $\mu_L$ passes through $U + \varepsilon_{1}$, 
see Fig.~7, there is a discontinuity in the current-voltage curve.
In the first regime,  the total current to the leading
approximation in $g_L, g_R$ is a sum
of two contributions, $I\simeq I_{eq} + I_{ne}$.
The first,  $I_{eq}=  e \Gamma_{0\to 0} + 2e \Gamma_{0\to 1}$, is
the equilibrium current coming from  cotunneling. 
The second contribution 
is associated with the nonequilibrium population of state ``1'' 
and is given by
\[
I_{ne} \!=\!2e\Gamma_{0\to 1}   \! \times \!\!\left\{ \begin{array}{ll}
\frac{\Gamma_{1 \to 0}-\Gamma_{0 \to 1} + 2 \Gamma_{1 \to 1}
- 2 \Gamma_{0 \to 0}}{ \Gamma_{0 \to 1}+ \Gamma_{1 \to 0}}
& \mu_L \! < \! U + \varepsilon_1 \\
2\left( 1+ \frac{ \Gamma_{e \to o}^{(1)}} { \Gamma_{e \to o}^{(0)}} \right) 
& \mu_L \!> \! U + \varepsilon_1 \end{array} \right. 
\]
Assuming that the  voltage drop $eV= \mu_L - \mu_R$ is larger than
the energy difference $\tilde{d}= \varepsilon_0 - \varepsilon_1$, 
the jump in the nonequilibrium current is:
\begin{equation}
\delta I_{ne}  =
 c_1 e\frac{ g_L g_R d^2}{ 8 \pi^2 h \tilde{d} }
\left( 1 - \frac{\tilde{d}}{eV} \right)\quad eV \sim 2\tilde{d} 
\label{ne-jump}
\end{equation}
where $c_1= 4 u_0^2 v_1^2 \rho_{L0} \rho_{R1}$
is a constant of order unity. This jump in the nonequilibrium current
leads to the peak in the differential conductance spectra.
Formula (\ref{ne-jump}) has simple interpretation.
Up to numerical prefactors it is a product of two factors: first is the 
probability of finding the grain with an unpaired electron in state 
``1''. It is proportional to $ g_R (d/\tilde{d})
(1-\tilde{d}/eV)$, and increases with
the voltage $V$ and as $\tilde{d}=\varepsilon_{0}-\varepsilon_1
\to 0$. The second factor is associated with the rate in 
which the state ``1'' is filled with an electron, $e g_L d/h$.  

\begin{figure}
\begin{center}
\leavevmode
\epsfxsize=8cm
\epsfbox{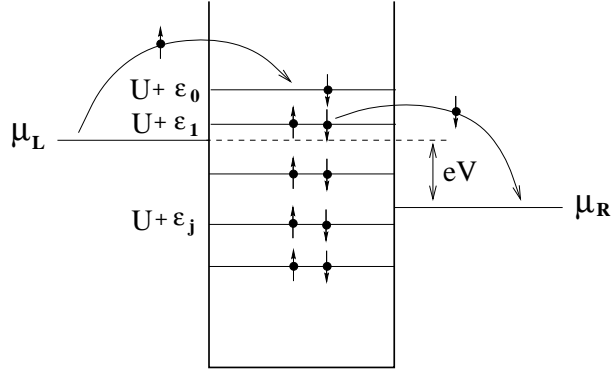} 
\end{center}
 \caption{Inelastic cotunneling process can drive an ``odd'' 
superconducting grain out of its ground state.
In the ground state, the single-particle level indicated by 
$U+\varepsilon_0$ is occupied by one electron. 
Excited states are those in which
the unpaired electron is shifted to other single-particle levels. 
In a nonequilibrium steady state, low single-particle levels become 
available for resonant tunneling, leading to a subresonances structure
of the differential conductance shown in Fig.~6. 
State $j$ shown to be filled with two electrons 
should be understood as a coherent superposition of double occupied and 
empty states with weights $v_j^2$ and $u_j^2$ respectively.}
\end{figure}

The magnitude of the jump (\ref{ne-jump}) should be compared to the
jump in the current as $\mu_L$ increases above $U+ \epsilon_0$, and
real transition via the even state become allowed.  
To the leading order in $g_L$ and $g_R$, the current
in this regime is 
\begin{equation}
I = c_2 e \frac{ g_R g_L d}{h (g_L + 4 g_R)}, \quad
\mu_L > U + \varepsilon_0,
\label{mainpeak}
\end{equation}
where $c_2$ is a constant of order unity having structure similar to
$c_1$.  Comparing the current jump, $\delta I_{ne}$, with that
associated with the resonant tunneling, $\delta I$, we find
\begin{eqnarray}
\frac{ \delta I_{ne}}{\delta I} \simeq
\frac{ g_L + 4 g_R}{8 \pi^2}, \quad 
eV \sim 2 \tilde{d}. \nonumber
\end{eqnarray} 
Thus nonequilibrium population of the excited level
of the odd-grain leads to the appearance of a subresonance at small $V$,
however, its height is much smaller than 
that of the main resonance.

We turn now to the second regime of the parameters, $d \ll eV <
\Delta$, in which many levels contribute to the transport. Again,
we focus our attention on the cotunneling regime, $\mu_L <
U+\varepsilon_0$.  We show that the characteristic amplitude
of the subresonances in this regime may become comparable to the 
amplitude of the main peak.
 
To the leading order in $g_L, g_R$, and $d/\Delta$, the steady 
state solution of the rate equations at $\mu_L= U+\varepsilon_1+0$ 
is $P_0 \simeq 1$, while for the other probabilities we have
\begin{equation}
P_e \simeq \frac{\sum_{i \neq 0 }\Gamma_{0\to i}}{\Gamma^{(0)}_{e\to o}},\quad
P_j \simeq 2 \frac{\Gamma_{0 \to j}}{\Gamma_{o \to e}^{(j)}}
+
2 \frac{\Gamma_{e \to o}^{(j)}}{\Gamma_{o\to e}^{(j)}}P_e.
\label{largeN} 
\end{equation}
The characteristic number of states contributing to the current
(\ref{current}) is
large as $\sqrt{\Delta eV}/d$ so that mesoscopic fluctuations of the
tunneling rates and of the inter-level spacings may be neglected. 
Additional large factor, $\sqrt{\Delta eV}/d$, comes from the summation
over the levels in Eq.~(\ref{largeN}), and we find
\begin{eqnarray}
I\simeq \frac{e^2 g_L g_R}{2\pi^2h}\frac{ V \Delta }
{\varepsilon_0-\varepsilon_j}, \quad
\left\{ \begin{array}{l} d \ll eV \ll \Delta \\ \mu_L = U+\varepsilon_j+0
\end{array} \right. . \nonumber
\end{eqnarray}
Once again, the current jumps  each time $\mu_L$ passes
through  $U +\epsilon_j$.
This jump for large $j$ (but still such that $U + \varepsilon_j 
-\mu_L \ll eV$) scales as $1/j^3$, and the ratio of the jump at $j=1$ 
to the jump at the resonance level (\ref{mainpeak}) is given by
\begin{eqnarray}
\frac{ \delta I_{ne}}{\delta I} \simeq
\frac{ (g_L + 4 g_R)}
{8 \pi^2} \frac{e V \Delta }{\tilde{d} d}, \quad
d \ll eV < \Delta. \nonumber
\end{eqnarray} 
Noticing that $\tilde{d}=\varepsilon_0-\varepsilon_1 \simeq d^2/ 2\Delta$,
we see that the first subresonance becomes comparable in height 
to the main one
at voltages as small as $eV \approx 4 \pi^2 d^3/ \Delta^2 ( g_L
+ 4 g_R)$.

We conclude by comparing the above results with the experimental data 
of Ref.~[\cite{Ralph1}]. There $\Delta
\approx 5 d$, the conductances in the normal state are $g_R \approx 10
g_L \approx 1/8$, and the leads are also superconducting. The
singularity in the density of states of the leads imply that the
effective conductance is increased by factor of $2-3$. 
Neglecting inelastic processes and the Josephson coupling, these
parameters imply that when $eV \approx 2 d$ the ratio of
the subresonances amplitude to that of the first resonance 
is of order one, while at $eV \sim 2 \tilde{d} \approx d/5$ it is of 
order of 1\%. It implies that first subresonance peak associated with
tunneling into state ``1'' cannot be resolved, both because its amplitude
and its distance from the main peak are too small. However next 
subresonance appear already at distance of order $d$ from the main
resonance, and for $V > 2d$ have an amplitude comparable with the  
main resonance.

\section{4~~Relaxation processes}

Central to our analysis is the assumption that the steady-state 
occupation configurations of the electrons in the dot are far 
from equilibrium. This condition holds when the rate  $1/\tau_{in}$ of 
inelastic relaxation processes is smaller than the tunneling 
rate of an electron into and out of the dot, $1/\tau_{tun}$. 
In the opposite limit, $1/\tau_{in} > 1/\tau_{tun}$, the system 
relaxes to equilibrium between 
tunneling events, and the electrons effectively occupy only one configuration. 
In this case one expects each resonance cluster to collapse 
to a single peak. This behavior is illustrated by the dashed line
in Fig.~5 where a large inelastic relaxation rate $1/\tau_{in}
=5/\tau_{tun}$ was included in the detailed-balance equations.

The data shown in Figs.~2 and 3 indicate that the metal particle in the
experimental system is indeed in a strongly nonequilibrium state. 
It is useful, however, to consider the various relaxation processes 
in our system in order to delimit the expected nonequilibrium regime.
Relaxation of excited Hartree--Fock states may occur due to: (1) 
electron-electron interaction in the dot beyond Hartree--Fock; 
(2) electron-phonon interaction; 
(3) Auger processes in which an electron in the dot relaxes 
while another one in the lead is excited; (4) relaxation of an 
electron in the dot as another electron tunnels out to the lead; (5) 
thermalization with the leads via tunneling. The last two processes 
are small corrections since they clearly happen on time scales larger than 
the tunneling time. 

In Ref.~[\cite{Altshuler96}] it was shown that excited  
many-body states of closed systems with energy $\epsilon$ smaller than 
$(g/\log g)^{1/2} d$ are merely slightly perturbed Hartree--Fock 
states. In other words, the overlap between the true many-body 
state and the corresponding Hartree--Fock approximation is
very close to unity. This justifies the use of our model for the low
energy resonances since $g \approx 5$  therefore the energy interval 
$0\! <\! \epsilon \! < \! (g/\log g)^{1/2}d$ contains at least
the first few excited states. 
At high source--drain voltage, however, when the dot is excited to
energy  $ g^{1/2}d\! <\! \epsilon \! <\! g d$, 
tunneling takes place into quasiparticle states of width 
$\epsilon^2/(g^2 d)$\refnote{\cite{Sivan94a}}. 
This width is larger than the typical separation between 
nearby resonances but smaller than $d$.
Therefore, electron-electron scattering will obliterate 
the fine structure of resonances for high energy excitations of the dot.

Consider now the electron--phonon interaction. 
The temperature, 30 mK, is much smaller than the
mean level spacing, therefore, the probability of phonon
absorption is negligible, and only emission may take place. 
The sound velocity in aluminum is $v_s=6420$ m/sec, 
therefore the wavelength of a phonon associated with relaxation
of energy $\omega \sim d =1$ meV is approximately 50 \AA, 
the same  as the system size. In this regime, 
we estimate the phonon emission rate to be  

\begin{eqnarray}
\frac{1}{\tau_{e-ph}} \sim 
\left(\frac{2}{3}\epsilon_F \right)^2 \frac{\omega^3 \tau d}
{2\rho \hbar^4 v_s^5}, \nonumber
\end{eqnarray}
where $\epsilon_F$ is the Fermi energy (11.7 eV in Al), and $\rho$ is 
the ion mass density (2.7 g/cm$^3$ in Al). This rate is that
of a clean metal but reduced by a factor of
$\tau d/\hbar$ where $\tau$ is the elastic mean free 
time\refnote{\cite{Reizer86}}. 
In ballistic systems, $\tau$ is the traversal time 
across the system of an electron at the Fermi level. Assuming  
ballistic motion this factor is of order 10$^{-3}$. 
The resulting relaxation rate for $\omega=d$ is therefore of order 
$1/\tau_{e-ph} \approx 10^8$ sec$^{-1}$ which is similar to the
tunneling rate $1/\tau_{tun}\approx 6 \cdot 10^8$ sec$^{-1}$ 
(corresponding to a current of $10^{-10}$ A through the particle). 
Thus, by increasing the resistance of the tunnel junctions one 
should be able to cross over to the
near-equilibrium regime shown by the dashed line in Fig.~5.

Relaxation due to Auger process is estimated to be negligible.
Two factors reduce this rate considerably: (1) it is exponentially 
small in $w/\chi$ where $w$ is the width of the tunnel junction 
and $\chi$ is the screening length; (2) interaction between 
electrons on both sides of the 
tunnel junction can take place only within a very limited volume.  

\section{5~~Conclusions}

In this review it was shown that the low-voltage tunneling-resonance 
spectra of a ultrasmall metallic grains, normal as well as  superconducting,
reflect nonequilibrium electron configurations.
These configurations are reached by resonant tunneling as well
as inelastic cotunneling. The first tunneling resonance develops 
a substructure on energy scales
of order of the single-particle mean level spacing, $d$, while high
resonances split due to electron-electron interactions and
appear in clusters of width $d/g$.  The latter phenomenon
is a result of electron-electron interaction beyond the 
orthodox model\refnote{\cite{Averin91}}.
Relaxation due to electron--phonon interaction,
which becomes important for high resistance tunnel barriers, 
will collapse the clusters. This effect can be used to probe
the electron-phonon relaxation rate in nanometer size metal
particles.

\subsection{Acknowledgments}
This review summarize results of collaboration with 
I.~L.~Aleiner, B.~L.~Altshuler, D.~C.~Ralph, M.~Tinkham, 
and N.~S.~Wingreen, whom it is my pleasure to thank. I would like
also to thank N.~Brenner for many useful comments on the manuscript. 

\section{Appendix A}
The purpose of this appendix is to calculate the second moment of
off-diagonal matrix elements of the interaction potential 
$U({\bf r}-{\bf r'})$, and show that
$U_{ijkl}$ are small as $1/g$, where $g$ is the dimensionless
conductance. The subject was discussed in several 
papers,\refnote{\cite{Blanter96,BM,AG}} 
and is presented here for completeness. 

When calculating off-diagonal matrix elements of the
interaction potential, it is important take into account
screening effects. The relevant two-particle
interaction potential is not the bare one, $\tilde{U}({\bf k})$,
but rather the statically screened potential: $\tilde{U}_s({\bf k}) =
\tilde{U} ({\bf k})/[ 1+ 2 \nu \tilde{U} ({\bf k}]$ where  
$\tilde{U}({\bf k})$ is the Fourier transform of the bare 
two-particle interaction $U({\bf r}-{\bf r'})$, and $\nu$ is the density
of states per unit volume. The contribution
to the off-diagonal matrix elements comes only from spatial 
fluctuations in the electron density (non-zero modes)
for which screening is established at very short time, 
of order of the time it takes for a plasmon to propagate through 
the system. The latter is much shorter than the relaxation
time, $t_c$, of fluctuations in the electron density. 
Thus for large $\nu$ the screened interaction potential, 
$U_s({\bf r}-{\bf r'})$, is close to a $\delta$-function. 

Consider, therefore, the integral
\begin{equation}
U_{ijkl} = \frac{1}{2 \nu} \int d{\bf r}  \psi_i^*({\bf r})
\psi_j^*({\bf r})\psi_l({\bf r})\psi_k({\bf r}), \label{u}
\end{equation}
where no two indices are the same. Clearly on average
$\langle U_{ijkl} \rangle =0$ since wave functions associated
with different eigenvalues are independent and $\langle \psi \rangle = 0$. 
To estimate the magnitude of the off-diagonal matrix elements we 
calculate the second moment $\langle |U_{ijkl}|^2 \rangle$. The square of
the matrix element, $|U_{ijkl}|^2$, contains four pairs of
wave functions in the form $\psi_i^*({\bf r})\psi_i({\bf r'})$.
Since the correlation between wave functions and eigenenergies
are only to order $1/g$, one can approximate these pairs as 
\begin{equation}
\psi_i^*({\bf r})\psi_i({\bf r'}) \approx \frac{d}{2 \pi i}
\left[ G({\bf r},{\bf r'};E_i-i\eta)- G({\bf r},{\bf r'};E_i + i\eta)
\right], \label{psi}
\end{equation}
where $d$ is the single-particle mean level spacing,
$\eta$ is a positive energy which will be taken to 
zero at the end of the calculation, and 
$G({\bf r},{\bf r'};E)$ is the single-particle Green function 
at energy $E$. Two basic correlators  emerge when calculating
the  ensemble or the energy average of $|U_{ijkl}|^2$.
These correlators, known in disordered diagrammatic nomenclature
as the diffuson and Cooperon, are
\begin{eqnarray}
\Pi_{\omega}({\bf r},{\bf r'})= \langle G({\bf r},{\bf r'};E+\omega+i\eta)
G({\bf r'},{\bf r};E - i\eta) \rangle, \nonumber
\end{eqnarray}
and $\langle G({\bf r},{\bf r'};E+i\eta) 
G({\bf r'},{\bf r};E+\omega - i\eta) \rangle$.
For systems with time reversal symmetry, considered here,
these correlators are the same. In the semiclassical limit,
\begin{equation}
\Pi_{\omega}({\bf r},{\bf r'})
= 2 \pi \nu \sum_{\mu } \frac{\bar{\chi}_{\mu}({\bf r}) 
\chi_{\mu}({\bf r'})}{-i\omega  + \hbar \gamma_\mu}, \label{pi}  
\end{equation}
where the sum is over all classical relaxation modes, i.e.
diffusion modes in the case of disordered grains and Perron-Frobenius 
modes in chaotic systems\footnote{For simplicity we consider here
chaotic systems in the form of billiards, namely the Hamiltonian
contains only a kinetic part, and chaotic dynamics is due to
the irregular boundary.}. $\gamma_\mu$ are the corresponding
eigenvalues, and $\bar{\chi}_{\mu}({\bf r})$  [$\chi_{\mu}({\bf r'})$] is the 
projection of the Perron-Frobenius left [right] eigenfunctions  
on the real coordinate space at fixed energy $E$. 

With the help of (\ref{u}), (\ref{psi}) and (\ref{pi}), and assuming
all energy differences (such as $E_i-E_j$) to be much smaller than
$\hbar \gamma_1$, one obtains
\begin{eqnarray}
\langle |U_{ijkl}|^2 \rangle \simeq c' \left( \frac{d}{g} \right)^2 \nonumber 
\end{eqnarray}
where $g$= $\hbar \gamma_1 / d$ is the dimensionless conductance of
the system, and $c'$ is a constant of order unity given by 
\begin{eqnarray}
c'= \frac{3 } {4 \pi^2}  \sum_{\mu \neq 0,\upsilon \neq 0} \int d{\bf r}
d{\bf r'} \mbox{Re} \frac{\bar{\chi}_\mu({\bf r})\chi_\mu({\bf r'})}
{\gamma_\mu / \gamma_1}
\mbox{Re} \frac{\bar{\chi}_{\upsilon}({\bf r'})\chi_{\upsilon}({\bf r})}
{\gamma_\upsilon / \gamma_1}. \nonumber
\end{eqnarray}
Here we assumed for simplicity that $\gamma_1$ is real. In case it 
contains also an imaginary part, the  same formula applies with the 
substitution $\gamma_1 \to \mbox{Re} \{ \gamma_1 \}$. 
Notice that there is no zero mode contribution to 
$\langle |U_{ijkl}|^2 \rangle$, since 
only density fluctuations associated with non zero-modes 
can induce scattering and contribute 
to $U_{ijkl}$. Mathematically this results from
the fact that eigenfunctions, $\chi_0({\bf r})$ and $\bar{\chi}_0({\bf r})$,
associated with the zero mode are real,
and since $\Pi_{\omega}$ is always calculated at a finite energy 
deference, $\omega$, taking its real part excludes the zero-mode 
contribution. 

The rest of this appendix is a semiclassical derivation of 
formula (\ref{pi}). We begin
by writing Green's function in the semiclassical approximation
as a sum of two terms\refnote{\cite{Gutzwiller}}:
\begin{eqnarray}
G({\bf r},{\bf r'};E \pm i\eta) \simeq G_0({\bf r},{\bf r'};E \pm i\eta)
+\sqrt{\frac{2 \pi}{h^f\hbar}}\sum_{l} A_{{\bf r}{\bf r'},l} 
e^{\pm \frac{i}{\hbar} S_{{\bf r}{\bf r'},l}(E)}. \nonumber
\end{eqnarray}
Here $G_0$ is the Weyl contribution associated with ``zero length''
trajectories. This term is important
only at distances $|{\bf r}-{\bf r'}|$ of order of the
particle wavelength and therefore can be neglected.
The second term is a sum over all classical trajectories
from  ${\bf r'}$ to ${\bf r}$, in which
$f$ is the number of degrees of freedom,
$S_{{\bf r}{\bf r'},l}(E)$ is the action, and 
$A_{{\bf r}{\bf r'},l}$ is the corresponding probability 
amplitude. It is convenient to introduce a local 
coordinate system in which $\tau$
is the time along the trajectory, and ${\bf r}_\perp$
are the coordinates perpendicular to the trajectory. In these
coordinates\refnote{\cite{Gutzwiller}}
\begin{eqnarray} 
|A_{{\bf r}{\bf r'},l}|^2 = \frac{1}{\dot{r} \dot{r}' }
\mbox{Det}^{-1} \left( \frac{\partial {\bf r}_{\perp}}
{\partial {\bf p'}_\perp} \right)_l, \nonumber
\end{eqnarray}
where $\dot{r}$ and $\dot{r}'$ denote  the velocity
of the particle at the final and initial points, and ${\bf p'}_\perp$ is the 
conjugate momenta to ${\bf r'}_\perp$. 

Expanding $S_{{\bf r}{\bf r'},l}(E+\omega)$ to first order in
$\omega$, and using the diagonal approximation for the product of 
the two Green functions, one obtains
\begin{equation}
\Pi_{\omega}({\bf r},{\bf r'}) 
= \frac{2 \pi}{h^f\hbar} \sum_{l} |A_{{\bf r}{\bf r'},l}|^2 
e^{i \omega T_l/\hbar}
= \frac{2 \pi}{h^f\hbar} \int dt P(t) e^{i \omega t/\hbar}, \label{midpi}
\end{equation}
where $T_l = \partial S_{{\bf r}{\bf r'},l}(E)/\partial E$ is the time
associated with the $l$-th trajectory, and
\begin{equation}
P(t) =  \sum_{l} \frac{1}{\dot{r} \dot{r}' }
\mbox{Det}^{-1} \left( \frac{\partial {\bf r}_{\perp}}
{\partial {\bf p'}_\perp} \right)_l \delta(t- T_l). \label{pt}
\end{equation}
Next we show that $P(t)$ is the projection of the classical propagator
in phase space onto configuration space at fixed energy $E$, namely
\begin{equation}
P(t) = \int d{\bf p'} \int d {\bf p}~ \delta[E- H({\bf r},{\bf p})]
~\langle {\bf r, p} | e ^{- {\cal L}t} | {\bf r',p'} \rangle, \label{ftpt}
\end{equation}
where $H({\bf r},{\bf p})$ is the classical Hamiltonian of the system
and $e ^{- {\cal L}t}$ is the evolution (Perron-Frobenius) operator 
for time $t$. In the coordinate system introduced above, 
the Hamiltonian function $H$ is the conjugate momentum to the time 
coordinate $\tau$ along the trajectory, therefore
\begin{eqnarray}
P(t) &=& \int \frac{d H}{\dot{r}} d{\bf p}_{\perp}
\int \frac{d H'}{\dot{r}'} d{\bf p'}_{\perp}
\delta (E-H) \delta(H- H'_t)\delta (\tau-\tau'_t) \delta({\bf r}_\perp-
{\bf r'}_{\perp t}) \delta({\bf p}_\perp- {\bf p'}_{\perp t}) \nonumber \\
&=&\frac{1}{\dot{r}\dot{r}'} \sum_l \delta(t-T_l) \int_{\Gamma_l} d 
{\bf p'}_{\perp}  \delta({\bf r}_\perp-{\bf r'}_{\perp t}), \nonumber 
\end{eqnarray}
where subscript $t$ denotes the value of the corresponding coordinate
after time $t$ starting from the phase space point $({\bf r',p'})$. 
Since the energy of the particle is fixed, the integral
reduces to a discrete sum over trajectories from ${\bf r'}$ to ${\bf r}$. 
The contribution to each trajectory,
$l$, comes from an infinitesimally small region of the coordinate
${\bf p'}_{\perp}$ denoted by $\Gamma_l$. Straightforward integration 
yields the result (\ref{pt}).  

Starting now form (\ref{ftpt}) and using the spectral decomposition of
the Perron-Frobenius operator
\begin{eqnarray}
\langle {\bf r, p} | e ^{- {\cal L}t} | {\bf r',p'} 
\rangle = f \delta( p^f-p'^f)
\sum_{\mu} e^{-\gamma_\mu t} \bar{\varphi}_\mu({\bf r,n})
\varphi_\mu({\bf r',n'}), \nonumber
\end{eqnarray}
where ${\bf n}$ denote the direction of the momentum, while $\bar{\varphi}_\mu$
and $\varphi_\mu$ are the left and right eigenfunctions, we get
\begin{eqnarray}
P(t) &=& h^f \nu \sum_{\mu} e ^{- \gamma_\mu t}  \int
\frac{d {\bf n} d{\bf n'}}{\Omega_f} 
\bar{\varphi}_\mu({\bf r,n})\varphi_\mu({\bf r',n'}) 
= h^f  \nu \sum_{\mu} e ^{- \gamma_\mu t} \bar{\chi}_{\mu}({\bf r}) 
\chi_{\mu}({\bf r'}). \nonumber
\end{eqnarray}
Here $\Omega_f = \int d{\bf n}$ is the solid angle of a sphere in 
$f$ dimensions, and 
$\nu = \int d{\bf p} \delta(E-H) /h^f$ defines the density of 
states per unit volume. Substituting this expression in (\ref{midpi})
and performing the time integration yields the required result (\ref{pi}).

\section{Appendix B}
In this appendix we derive formula (\ref{epsilon}) for the
cost of introducing an additional electron into an odd state:
$\varepsilon_j= E_{2m}- E_{2m-1}^{(j)}$ (ignoring the charging energy).
Denote by $F_N = E_{N} - \mu_N N$ the free energy of the system where
$E_N$ is the ground state energy with $N$ electrons, and $\mu_N$ is the 
chemical potential. 
For the superconducting state with unpaired electron in level
$j$ one has 
\begin{equation}
F_{2m-1}^{(j)} = \sum_{k\neq j} (\xi_k - 
\epsilon_k) + \Delta^2/\lambda + \xi_j, ~~~~~~~~~~~
\epsilon_k=\sqrt{\xi_k^2 + \Delta^2},
\end{equation}
where $\lambda$ is the pairing coupling constant.
In the intermediate state  where another electron 
tunneled into the $j$-th state and pairs with the already existing
one: $F_{2m} =  \sum_{k} (\xi_k -\epsilon_k) + \Delta^2/\lambda $. 
The parameters in these two equations, $\Delta$ and $\mu$ 
(which are functions of $N$),
are determined by the relations:
\begin{eqnarray}
\frac{\partial F_{N}}{\partial \Delta_{N}} = 0 ~~~~~~~~~~~~~~~~~~~~
N = - \frac{\partial F_{N}}{\partial \mu_{N}} \nonumber
\end{eqnarray}
Notice that the single-particle energies $\xi_k$ are measured with
respect to the chemical potential, thus for the differentiation with
respect to $\mu_N$ it it is convenient to introduce
$\xi_k= \tilde{\xi}_k- \mu_{N}$, and
$\epsilon_k= \sqrt{ (\tilde{\xi}_k- \mu_{N})^2 + \Delta_{N}^2}$,
where $\tilde{\xi}_k$ are independent of $\mu_N$. 

The above derivatives for the {\em even case}, $N=2m$, give
\begin{equation}
 \sum_k \frac{1}{2\sqrt{ (\tilde{\xi_k} - \mu_{2m})^2 + \Delta_{2m}^2}} 
= \frac{1}{\lambda}, ~~~~~~~~~
2m = \sum_{k} \left( 1 - \frac{\tilde{\xi_k} - \mu_{2m}}{
\sqrt{ (\tilde{\xi_k} - \mu_{2m})^2 + \Delta_{2m}^2}} \right), \label{even}
\end{equation}
while for the {\em odd case}, $N=2m-1$,
\begin{equation}
 \sum_{k \neq j} 
\frac{1}{2\sqrt{ (\tilde{\xi_k} - \mu_{2m-1}^{(j)})^2 + \Delta_{2m-1}^2}} 
= \frac{1}{\lambda}, ~~~~~~
2m-2 = \sum_{k\neq j} \left( 1 - \frac{\tilde{\xi_k} - \mu_{2m-1}^{(j)}}{
\sqrt{ (\tilde{\xi_k} - 
\mu_{2m-1}^{(j)})^2 + \Delta_{2m-1}^2}} \right). \label{odd}
\end{equation}
We expand $\varepsilon_j= E_{2m}-E_{2m-1}^{(j)} =
F_{2m}(\mu_{2m}; \Delta_{2m}) - F_{2m-1}^{(j)}(\mu_{2m-1};\Delta_{2m-1}) 
+ 2m \mu_{2m} - (2m-1) \mu_{2m-1}$ to linear order in the differences 
$\Delta_{2m}- \Delta_{2m-1}$, and $\mu_{2m}-\mu_{2m-1}$: 
\begin{eqnarray} 
\varepsilon_j & \simeq & F_{2m}(\mu_{2m}; \Delta_{2m}) - 
F_{2m-1}(\mu_{2m};\Delta_{2m})+  \mu_{2m} \nonumber \\  
& = & \mu_{2m-1}^{(j)} - \sqrt{ (\tilde{\xi}_j - \mu_{2m})^2 + \Delta_{2m}^2}. 
 \label{vej}
\end{eqnarray}
We are left, now, with the problem of finding the dependence
of $\mu_{2m-1}^{(j)}$ on the unpaired electron state $j$.
For this purpose we choose $\mu_{2m}$ as the reference point and
calculate the difference $\mu_{2m-1}^{(j)}- \mu_{2m}$.
From Eqs. (\ref{even}) and (\ref{odd}) we have
\begin{eqnarray} 
2m-2 &=&  \sum_{k \neq j} \left( 1 - \frac{\tilde{\xi_k} - \mu_{2m-1}^{(j)}}{
\sqrt{ (\tilde{\xi_k} - \mu_{2m-1}^{(j)})^2 + \Delta_{2m-1}^2}} \right) 
\nonumber \\
&=& \sum_{k} \left( 1 - \frac{\tilde{\xi_k} - \mu_{2m}}{
\sqrt{ (\tilde{\xi_k} - \mu_{2m})^2 + \Delta_{2m}^2}} \right) -
 \left( 1 - \frac{\tilde{\xi_j} - \mu_{2m}}{
\sqrt{ (\tilde{\xi_j} - \mu_{2m})^2 + \Delta_{2m}^2}} \right) \nonumber \\
  &-& ( \mu_{2m-1}^{(j)} - \mu_{2m} ) \sum_{k} \left(  \frac{-1}{
\sqrt{ (\tilde{\xi_k} - \mu_{2m})^2 + \Delta_{2m}^2}} 
+  \frac{(\tilde{\xi_k} - \mu_{2m})^2}{
\left( (\tilde{\xi_k} - \mu_{2m})^2 + \Delta_{2m}^2 \right)^{3/2} } 
\right). \nonumber
\end{eqnarray}
(The correction to $\Delta$ vanishes in the limit where the single-particle
mean level spacing $d$ is much smaller than the superconducting gap,
$d \ll \Delta$). 
Since the first sum in the second line equals $2m$ we have
\begin{eqnarray} 
-2 &=& 1 - \frac{\tilde{\xi_j} - \mu_{2m}}{
\sqrt{ (\tilde{\xi_j} - \mu_{2m})^2 + \Delta_{2m}^2}} 
+ \sum_k  \frac{ (\mu_{2m-1}^{(j)} - \mu_{2m})\Delta_{2m}^2}{
\left( (\tilde{\xi_k} - \mu_{2m})^2 + \Delta_{2m}^2 \right)^{3/2} } 
\nonumber \\
&\simeq &  1 - \frac{\tilde{\xi_j} - \mu_{2m}}{\Delta}
+ (\mu_{2m-1}^{(j)} - \mu_{2m}) \int \frac{d \xi}{d} 
\frac{\Delta^2}{\left(\xi^2 + \Delta^2 \right)^{3/2}}. \nonumber
\end{eqnarray}
Thus
\begin{eqnarray} 
\mu_{2m} \simeq \mu_{2m-1}^{(j)} +  \frac{d}{2} \left(3 
- \frac{\xi_j}{\Delta} \right),  \nonumber
\end{eqnarray}
and substituting this result in (\ref{vej}) we obtain (\ref{epsilon}).

\end{document}